\documentclass[aps,prl,twocolumn,toolkits,superscriptaddress,nofootinbib]{revtex4}

\usepackage{graphicx,bm} \usepackage{amsmath,amssymb}
\usepackage{mathrsfs,verbatim,subfigure}

\renewcommand\section[1]{{\em #1.}}

\begin{document}

\title{Universal Parametrization of Thermal Photon Rates in Hadronic Matter}

\author{Matthew Heffernan}
\email{mrheffernan@email.wm.edu}
\affiliation{Cyclotron Institute and Department of Physics and Astronomy, Texas A\&M University,
College Station, TX 77843-3366, USA}
\affiliation{Department of Physics, College of William \& Mary, Williamsburg, VA 23187-8795, USA}

\author{Paul Hohler}
\email{pmhohler@comp.tamu.edu}
\affiliation{Cyclotron Institute and Department of Physics and Astronomy, Texas A\&M University,
College Station, TX 77843-3366, USA}

\author{Ralf Rapp}
\email{rapp@comp.tamu.edu}
\affiliation{Cyclotron Institute and Department of Physics and Astronomy, Texas A\&M University,
College Station, TX 77843-3366, USA}

\begin{abstract}
Electromagnetic (EM) radiation off strongly interacting matter created in high-energy
heavy-ion collisions (HICs) encodes information on the high-temperature phases of nuclear
matter.  Microscopic calculations of thermal EM emission rates are usually rather
involved and not readily accessible to broad applications in models of the
fireball evolution which are required to compare to experimental data. An accurate and
universal parametrization of the microscopic calculations is thus key to honing the
theory behind the EM spectra. Here we provide such a parametrization for photon
emission rates from hadronic matter, including the contributions from in-medium
$\rho$ mesons (which incorporate effects from anti-/baryons), as well as
Bremsstrahlung from $\pi\pi$ scattering. Individual parametrizations for each
contribution are numerically determined through nested fitting functions for photon
energies from 0.2 to 5\,GeV in chemically equilibrated matter of temperatures
100-180\,MeV and baryon chemical potentials 0-400\,MeV. Special care is taken to
extent the parameterizations to chemical off-equilibrium as encountered
in HICs after chemical freezeout. This provides a functional description of thermal
photon rates within a 20\% variation of the microscopically calculated values.
\end{abstract}

\maketitle


\section{Introduction}
The understanding of hot and dense QCD matter remains a primary goal in nuclear physics.
This is experimentally pursued through ultra-relativistic heavy-ion collisions (URHICs),
producing a fireball of strongly interacting matter which expands and cools. Photons
are an interesting probe of this fireball because they are emitted throughout its
lifetime and reach the detectors without further interactions with the medium,
see~Refs.~\cite{Rapp:2009yu,Gale:2009gc} for recent reviews. The experimentally
measured spectra depend on both the microscopic production mechanisms and the bulk
evolution of the fireball. Recent measurements of direct-photon spectra and their
elliptic flow~\cite{Adare:2008fq,Adare:2011zr,Wilde:2012wc,Lohner:2012ct,Yang:2014mla}
have triggered intense activity to understand the data~\cite{Liu:2009kta,vanHees:2011vb,
Holopainen:2011pd,Dion:2011pp,Mohanty:2011fp,Shen:2013vja,Linnyk:2013wma,Bzdak:2012fr,
Monnai:2014kqa,Gale:2014dfa,McLerran:2014oea}.
In particular, the magnitude of the elliptic flow points to large contributions from
intermediate and late phases of the fireball evolution, i.e., the pseudo-critical region
and the hadronic phase~\cite{vanHees:2011vb}.

In the present paper, we focus on thermal emission from the hot and dense hadronic
medium. Early works on this problem have concentrated on hot meson
matter~\cite{Kapusta:1991qp,Xiong:1992ui,Steele:1996su,Sarkar:1997aa}; a rather
detailed analysis of the $\pi\rho a_1$ system (with an extension to strangeness) has been
conducted in Ref.~\cite{Turbide:2003si}, where pertinent rate parameterizations
in photon energy ($q_0$) and temperature ($T$) have also been given. Based on developments
in the dilepton sector~\cite{Rapp:1997fs,Urban:1999im}, it was realized that
baryonic emission sources play an important role for photon rates, by carrying the
in-medium $\rho$ spectral function to the photon point~\cite{Turbide:2003si}. The
underlying many-body calculations of the $\rho$ spectral function, which account
for pion-cloud modifications (corresponding to pion exchange reactions,
including Bremsstrahlung) and resonant $\rho$-hadron interactions (corresponding to
resonance Dalitz decays)~\cite{Urban:1999im,Rapp:1999us,Rapp:1999qu}, are rather
involved and as such not readily available for a broad use in evolution models of
URHICs.
The main objective of the present paper is to provide compact parameterizations of
these photon rates which for the first time encompass a finite baryon chemical
potential ($\mu_B$) as an additional variable. We also revisit the problem of hadronic
Bremsstrahlung, specifically for the most abundant $\pi\pi\to\pi\pi\gamma$ channel,
by extending the calculations of Ref.~\cite{Liu:2007zzw} to higher energies and
providing pertinent parameterizations as well.

\section{Thermal photon rate parametrizations}
We first consider thermal photons emitted from in-medium $\rho$ mesons;
the pertinent rates can be cast in terms of the transverse
electromagnetic (EM) spectral
function, $\rho_{\rm EM}^T$, as~\cite{McLerran:1984ay}
\begin{equation}
q_{0}\frac{dR_{\gamma}}{d^{3}q}\left(q_0;\mu_B,T\right)=
\frac{\alpha_{\rm EM}}{\pi}f^{B}(q_{0};T)
\rho^{T}_{\rm EM}(q_{0}=q;\mu_B,T)
\end{equation}
with $f^{B}(q_{0};T)=1/[e^{(q_{0}/T)}-1]$: Bose distribution function, and
$\alpha_{\rm EM}$=1/137. By invoking (a generalized) vector meson dominance,
the EM spectral functions can be related to the in-medium $\rho$
propagator. The latter has been developed in
Refs.~\cite{Rapp:1997fs,Urban:1999im,Rapp:1999qu,Rapp:1999us} and leads to a
strong broadening of the spectral peak due to interactions with baryons and
anti-baryons, which are critical in describing experimental dilepton spectra
from URHICs~\cite{Rapp:2013nxa}. Photon rates are readily extracted
from the light-like limit of vanishing invariant mass, $M\to0$, and depend
on energy, $q_0$, temperature, $T$, and baryon chemical potential, $\mu_B$.

\begin{figure}[!t]
\centering
\includegraphics[width=.45\textwidth]{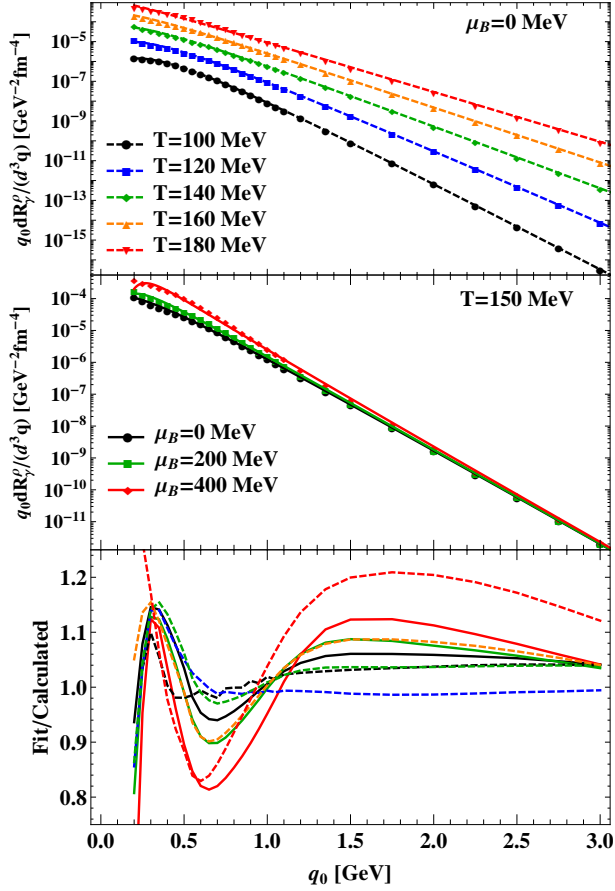}
\caption{Photon emission rates (vs.~photon energy) calculated from in-medium $\rho$
mesons (points) compared to their parametrization, Eq.~(\ref{eq:fact}) (curves), at
$\mu_B$=0 (upper panel) and $T$=150\,MeV (middle panel). Bottom panel:
Ratio of parametrization over calculated rates.}
\label{fig:ratecomp}
\end{figure}
In a first step of constructing a parametrization, the photon emission rates have
been explicitly calculated~\cite{Rapp:1999us} at $\mu_B$=0 for a set of 9
temperatures, $T$=100, 110, $\dots$, 180\,MeV, shown by the symbols in
Fig.~\ref{fig:ratecomp} (note that these rates include effects due to (equal
densities of) baryons and anti-baryons, whose contribution, however, is strongly
suppressed by the thermal weight as $T$ decreases). These rates are then
parameterized at each temperature by the ansatz
\begin{equation}
\label{eq:mu0}
q_{0}\frac{dR_{\gamma}^\rho}{d^{3}q}\left(q_0;0,T\right)
=\exp\left[a\left(T\right)q_0+b\left(T\right)
+\frac{c\left(T\right)}{q_0+0.2}\right] \ ,
\end{equation}
and smooth $T$ dependencies of the parameters are found as
\begin{equation}
\begin{split}
\label{eq:mu0b}
a\left(T\right) &= -31.21+353.61T-1739.4T^2+3105T^3, \\
b\left(T\right) &= -5.513-42.2T+333T^2-570T^3,\\
c\left(T\right) &= -6.153 + 57T -134.61T^2+8.31T^3 \ .
\end{split}
\end{equation}
In all parametrizations, $q_0$, $T$, and $\mu_B$ are in units of GeV.
The fit results are indicated by the symbols in the upper panel of
Fig.~\ref{fig:ratecomp}, and their ratio over the calculated results is
displayed in the bottom panel. For the highest temperature of $T$=180\,MeV,
the deviations can reach up to 20\%, but are within about 10\% for all
other temperatures and energies up to 5\,GeV (note that the blue-shift
due to the radial fireball expansion in URHICs implies an
appreciable shift of the restframe energy to the lab energy, e.g., by
about a factor of 2 in Au-Au($\sqrt{s}$=200\,GeV) at RHIC
energies~\cite{Rapp:2014qha}).

In a second step, the microscopic rates are calculated for three finite
baryon chemical potentials, $\mu_B$=0.1, 0.2 and 0.4\,GeV, and for each one
at the nine temperatures quoted above. Based on the $\mu_B$=0 fits above,
the following factorized ansatz was made
\begin{equation}
\label{eq:fact}
q_{0}\frac{dR_{\gamma}^\rho}{d^{3}q}\left(q_0;\mu_B,T\right) =
q_0\frac{dR_\gamma^\rho}{d^{3}q}\left(q_0;0,T\right) F^\rho\left(q_0;\mu_B,T\right)
\end{equation}
with the function
\begin{equation}
F^\rho\left(q_0;\mu_B,T\right) = \exp\left[d\left(\mu_B,T\right)-\frac{k\left(\mu_B,T\right)}{q_0^{2}}-\frac{m\left(\mu_B,T\right)}{q_0}\right] \, .
\end{equation}
The parameters $d$, $k$, and $n$ are determined from fits at fixed
$T$ to determine their $\mu_B$ dependence through an expansion as
\begin{eqnarray}
d\left(\mu_B,T\right)& =& n\left(T\right)\mu_B + p\left(T\right)\mu_B^2 + r\left(T\right)\mu_B^3, \nonumber\\
k\left(\mu_B,T\right)&=&s\left(T\right)\mu_B + v\left(T\right)\mu_B^2+w\left(T\right)\mu_B^3,\\
m\left(\mu_B,T\right)&=&\alpha\left(T\right)\mu_B+\beta\left(T\right)\mu_B^2+\eta\left(T\right)\mu_B^3 \, . \nonumber
\end{eqnarray}
Lastly, the $T$ dependence in the above coefficients fit via smooth
functional dependencies resulting in
\begin{eqnarray}
\label{eq:mupara}
n\left(T\right)&=&- 0.04+2.3T-12.8T^2, \nonumber\\
p\left(T\right)&=& 23.66-354T+1175T^2, \nonumber\\
r\left(T\right)&=&-54.3+742.6T-2350T^2, \nonumber\\
s\left(T\right)&=&-22.11+808.7T-11604.4T^2 \nonumber \\
&&+81700T^3-282480T^4+384116T^5, \nonumber\\
v\left(T\right)&=&-1.6-121.7T+1775T^2-5516T^3,\\
w\left(T\right)&=&-9.874+469T-4371.5T^2+11000T^3, \nonumber\\
\alpha\left(T\right)&=&84.784-3028.6T+42434T^2 \nonumber \\
&&-291390T^3+981000T^4-1295400T^5, \nonumber\\
\beta\left(T\right)&=&59.64-726.46T+1093.4T^2+4256T^3, \nonumber\\
\eta\left(T\right)&=&-73.9+458.66T+2450T^2-12348T^3\, . \nonumber
\end{eqnarray}
A comparison between this parametrization and the explicitly calculated
rates is shown in Fig.~\ref{fig:ratecomp}. We find that the parametrization
reproduces the calculated rates with an accuracy better than 20\%.

As a final test of the reliability of our fits, the ``predictions" from the
parametrization are compared to the calculated rates at $\mu_B=0.3$\,GeV,
a value not used in the fitting procedure. The deviation between
parametrization and calculation is found to be very similar to fitted
cases. Therefore, we conclude that our parameterized photon rates lie within
the 20\% error margin (significantly smaller for the most part) established
in the fits, for photon energies $q_0$=0.2-5\,GeV, temperatures
$T$=100-180\,MeV, and baryon chemical potentials $\mu_B$=0-0.4\,GeV.

Processes of type $\pi N\to\pi N\gamma$ and $NN\to NN\gamma$ are included in the
$\rho$ spectral functions used in the fits above, but meson-meson Bremsstrahlung
is not. Since pions are the most abundant mesons at the relevant temperatures,
and their small mass renders the kinematics favorable for radiating off photons,
the dominant source in the mesonic sector is expected from
$\pi\pi \rightarrow \pi\pi\gamma$ processes. The pertinent rates have been calculated
in Ref.~\cite{Liu:2007zzw} in an effective hadronic model for $S$- and $P$-wave
$\pi\pi$ (and $\pi K$) scattering. Special care was taken in maintaining EM gauge
invariance in the presence of hadronic form factors, and in going beyond the often
times applied soft-photon approximation. The analysis, however, focused on rather
small photon energies, below 0.5\,GeV, thus limiting the applicability of the
provided parameterizations. Here, we carry these calculations to higher energies
and generate suitable parameterizations.

\begin{figure}[!t]
  \centering
\includegraphics[width=.45\textwidth]{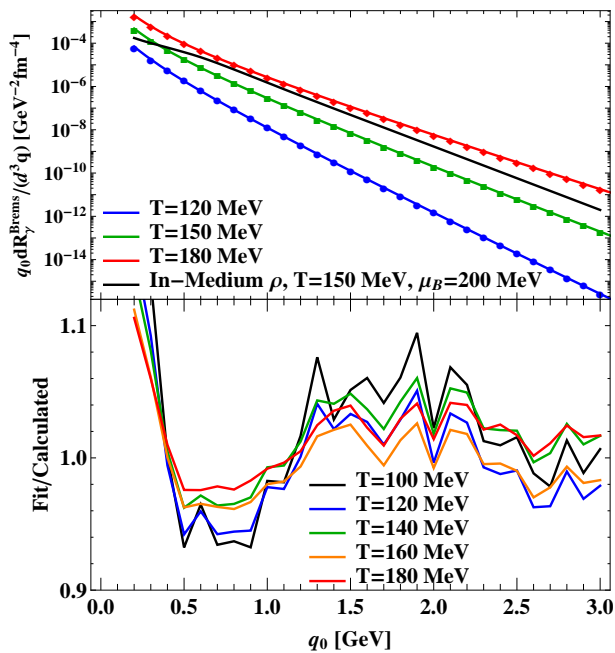}
\caption{Upper panel: Calculated thermal photon rates from $\pi\pi$ Bremsstrahlung (symbols)
compared to their parametrization (colored lines). As a reference, we also display a rate
from calculated in-medium $\rho$ decays (black line).
Lower panel: Ratio of the parameterized over calculated Bremsstrahlung rates.}
\label{fig:brems}
\end{figure}
Let us first compare the $\pi\pi$ Bremsstrahlung rate to the in-medium $\rho$
decays discussed above. At typical hadronic temperatures of $T$=150\,MeV
the former exceeds the latter for $q_0$$<$0.4\,GeV, but drops below it by about
an order of magnitude for $q_0$$\geq$1\,GeV, see Fig.~\ref{fig:brems}.
We note in passing that the contribution from  $\pi K$ Bremsstrahlung amounts
to about 20\% of the $\pi\pi$ one~\cite{Liu:2007zzw}. We have parameterized the latter
using the ansatz
\begin{equation}
\begin{split}
q_{0}\frac{dR_{\gamma}^{\rm Brems}}{d^{3}q}&\left(q_0; T\right)  = \exp\left[ \alpha_{B}(T) +
\beta_B(T) q_0 \right.
\\
&\left. + \gamma_B(T) q_0^2 +\delta_B(T) (q_0+0.2)^{-1}\right],
\end{split}
\label{brems}
\end{equation}
and found that with
\begin{equation}
\begin{split}
\alpha_B(T) &= -16.28 +62.45 T -93.4 T^2-7.5 T^3,  \\
\beta_B(T) &= -35.54 + 414.8 T -2054 T^2 +3718.8 T^3,  \\
\gamma_B(T) &= 0.7364 - 10.72 T +56.32 T^2-103.5 T^3, \\
\delta_B(T) &= -2.51 + 58.152 T -318.24 T^2 +610.7 T^3 \,,
\end{split}
\end{equation}
the calculated rates are fitted within $\sim$5\% for $T$=100-180\,MeV and $q_0$=1-5\,GeV,
cf.~lower panel of Fig.~\ref{fig:brems}\footnote{We note that when combining the present
$\pi\pi$ Bremsstrahlung rates with the ones given in the appendix of Ref.~\cite{Turbide:2003si},
the $\rho\to\pi\pi\gamma$ and $\pi\pi\to\rho\gamma$ contributions in there need to be dropped
to avoid double-counting.}. 
Noise is statistical in nature due to the calculated
rates.

\section{Chemical off-equilibrium (COE)}
\label{sec:evol}
The above para\-metri\-zations pertain to hadronic matter in chemical equilibrium (CE),
i.e., for $\mu_B = -\mu_{\bar{B}}$ without any other chemical potentials (and
therefore $\mu_B\equiv \mu_N$).
However, in URHICs, hadro-chemical freezeout occurs well before kinetic freezeout,
implying the emergence of effective chemical potentials, $\mu_i$,  to conserve the
ratios of hadrons which are stable under strong decay, e.g., $i$=$\pi$, $K$,
baryons and anti-baryons~\cite{Rapp:2002fc}. Since strong resonance formation
reactions persist, one has $\mu_\rho$=2$\mu_\pi$, $\mu_\Delta$=$\mu_N+\mu_\pi$,
etc. An extension of the rate para\-metri\-zations to fully incorporate the $\mu_i$
dependencies is not practical. However, their leading effect can be rather
accurately captured by fugacity factors. For $\pi\pi$ Bremsstrahlung, this
amounts to an extra overall factor of $z_\pi^2$ on the right-hand-side of
Eq.~(\ref{brems}), with $z_\pi$=$\exp(\mu_\pi/T)$. The same factor also applies to
Eq.~(\ref{eq:fact}) (representing the $\rho$ fugacity), but additional amendments
are needed, as we will discuss now.

Let us start from chemical freezeout, $T_{\rm ch}$, where
$\mu_{\bar{B}}^{\rm ch} = -\mu_B^{\rm ch}$. For $T<T_{\rm ch}$, the separate
conservation of baryon and anti-baryon number causes the effective anti-baryon
chemical potential to {\rm rise} with $\mu_B$ approximately as
$\mu_{\bar{B}}(T)=\mu_{\bar{B}}^{\rm ch}+(\mu_B(T)-\mu_{B}^{\rm ch})
= \mu_B(T)  -2 \mu_B^{\rm ch}$~\cite{Rapp:2002fc}.
This increase in $\mu_{\bar{B}}(T)$ over the CE case must be
accounted for in the baryonic contributions to the rate. Toward this end, we
define the ratio $r$ by which the COE density of baryons {\rm plus}
anti-baryons is enhanced over the CE value,
\begin{equation}
\label{eq:rvalue}
r \equiv \frac{n_{B+\bar{B}}^{\rm COE}}{n_{B+\bar{B}}^{\rm CE}} =
\frac{n_B(\mu_B) + n_{\bar B}(\mu_B-2\mu_B^{\rm ch})}{n_{B}(\mu_B)+n_{\bar B}(-\mu_B)}
= \frac{1+{\rm e}^{-2\mu_B^{\rm ch}/T}}{1+{\rm e}^{-2\mu_B/T}} \ .
\end{equation}
Here, we have utilized the Boltzmann approximation,
$n_B(\mu_B)\simeq n_B(0)\,{\rm e}^{\mu_B/T}$.
The effective baryon chemical potential, $\mu_B^{\rm eff}$, to be used in the
above photon rate, is then given by
\begin{equation}
\label{eq:mu_eff}
{\rm e}^{\mu_B^{\rm eff}/T} \equiv r~{\rm e}^{\mu_B/T} \ \
\Rightarrow \ \ \mu_B^{\rm eff} = \mu_B + T\log(r) \ .
\end{equation}

The thermal meson-induced photon emission in the $\rho$ spectral function
is mostly due to resonance formation, $P\rho\to M\to P\gamma$~\cite{Rapp:1999qu}.
With $P$=$\pi$ being the dominant contribution, one picks up another factor of $z_\pi$
(in addition to the $z_\pi^2$ of the $\rho$). The meson gas sources only prevail at
higher $q_0$, while the baryon-induced ones take over toward smaller $q_0$.
However, many of the baryons are in excited states which carry
larger chemical potentials than the nucleon, e.g.,  $\mu_\Delta$=$\mu_N+\mu_\pi$,
$\mu_{N(1520)}$=$\mu_N+1.45\mu_\pi$, etc. It turns out that an extra overall factor of
$z_\pi$ approximately accounts for the chemically enhanced resonance abundances.

\begin{figure}[t!]
  \centering
	\includegraphics[width=.45\textwidth]{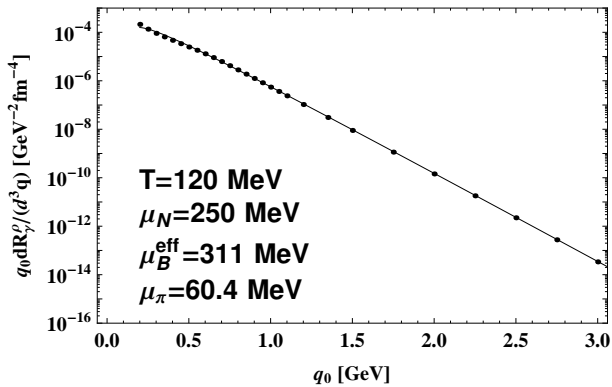}
\caption{Photon emission rates calculated from in-medium $\rho$'s (points)
compared to the parametrization of Eq.~(\ref{eq:fact}) evaluated with
$\mu_B^{\rm eff}$ from Eq.~(\ref{eq:mu_eff}) and an overall fugacity of $z_\pi^3$.}
\label{fig:off-eq}
\end{figure}

To summarize the overall effect of the COE extension, the function
$F^\rho$ in Eq.~(\ref{eq:fact}) should be replaced as
\begin{equation}
F^\rho \rightarrow z_{\pi}^3 \ F^\rho \left(q_0;\mu_B^{\rm eff},T\right) ,
\end{equation}
while the $\pi\pi$ Bremsstrahlung rate receives
an overall factor of $z_\pi^2$. These amendments yield rather accurate agreements,
typically within less than 10\% (see, e.g., Fig.~\ref{fig:off-eq}), largely
determined by the intrinsic uncertainty of the equilibrium parametrizations.

\section{Conclusion}
\label{sec:con}
We have constructed universal parametri\-zations for microscopic photon emission rates
from in-medium $\rho$ mesons and $\pi\pi$ Bremsstrahlung over a range
of photon energies, temperatures, and baryon-chemical potentials relevant to applications
in URHICs. Our para\-metrizations reproduce the calculated rates within 20\% (mostly
within 10\%). We have confirmed that $\pi\pi$ Bremsstrahlung is appreciable for energies
$q_0 <1$\,GeV, but subleading above. We have devised a prescription to extend the
equilibrium parameterizations to capture the effects of chemical off-equilibrium as
encountered in URHICs. We believe that these parametrizations will be useful in
calculations of thermal photon emission within different medium evolution models, and
thus contribute to a better understanding of pertinent observables.

\acknowledgments

This work has been supported by the U.S.~NSF under REU-grant No.~PHY-1263281 and
grant No.~PHY-1306359, and by the Humboldt Foundation (Germany).

\end{document}